\documentstyle[aps,prl,preprint,floats,epsfig]{revtex}

\begin{document}

\preprint{\tighten\vbox{\hbox{\hfil CLNS 00-1688}
                        \hbox{\hfil CLEO 00-17}
}}

\newcommand{\GG}{\mbox{$\gamma\gamma$}}
\newcommand{\PP}{\mbox{$\pi^+\pi^-$}}
\newcommand{\MM}{\mbox{$\mu^+\mu^-$}}
\newcommand{\LL}{\mbox{$\ell^+\ell^-$}}
\newcommand{\EE}{\mbox{$e^+e^-$}}
\newcommand{\DE}{\mbox{$\Delta E$}}
\newcommand{\BR}{\mbox{$\mathcal B$}}
\newcommand{\KPZ}{\mbox{$K^{*+} \to K^+ \pi^0$}}
\newcommand{\KSP}{\mbox{$K^{*+} \to K_S^0 \pi^+ \to \pi^+ \pi^- \pi^+$}}
\newcommand{\KSPP}{\mbox{$K^{*+} \to K_S^0 \pi^+$}}
\newcommand{\BMK}{\mbox{$B^+ \to \psi(2S) K^+$}}
\newcommand{\BMKS}{\mbox{$B^+ \to \psi(2S) K^{*+}$}}
\newcommand{\BMSS}{\mbox{$B^+ \to \psi(2S) K^{*+}$}}
\newcommand{\BZK}{\mbox{$B^0 \to \psi(2S) K^0$}}
\newcommand{\BZKS}{\mbox{$B^0 \to \psi(2S) K^{*0}$}}
\newcommand{\BZSS}{\mbox{$B^0 \to \psi(2S) K^{*0}$}}

\title{\large Study of $B \to \psi(2S) K$ and $B \to \psi(2S) K^*(892)$ decays}
 
\author{(CLEO Collaboration)}
\maketitle
\tighten

\begin{abstract}

   Color-suppressed decays of $B$ mesons to final 
states with $\psi(2S)$ mesons
have been observed with the CLEO detector. The branching fractions
for the decays $B^+ \to \psi(2S) K^+$, $B^+ \to \psi(2S) K^*(892)^+$,
$B^0 \to \psi(2S) K^0$, and $B^0 \to \psi(2S) K^*(892)^0$
are  measured to be  
$(7.8 \pm 0.7 \pm 0.9) \times 10^{-4}$,
$(9.2 \pm 1.9 \pm 1.2) \times 10^{-4}$,
$(5.0 \pm 1.1 \pm 0.6) \times 10^{-4}$, and
$(7.6 \pm 1.1 \pm 1.0) \times 10^{-4}$, respectively, where the
first uncertainty is statistical and the second is systematic.
The  first measurement of the longitudinal polarization fraction is 
extracted from the
angular analysis of the $B \to \psi(2S) K^*(892)$ candidates:
$\Gamma_L / \Gamma = 0.45 \pm 0.11 \pm 0.04 $.
Our measurements of the decays $B^0 \to \psi(2S) K^0$ and
$B^+ \to \psi(2S) K^*(892)^+$ are first observations. 

\end{abstract}

\newpage

\begin{center}
S.~J.~Richichi,$^{1}$ H.~Severini,$^{1}$ P.~Skubic,$^{1}$
A.~Undrus,$^{1}$
S.~Chen,$^{2}$ J.~Fast,$^{2}$ J.~W.~Hinson,$^{2}$ J.~Lee,$^{2}$
D.~H.~Miller,$^{2}$ E.~I.~Shibata,$^{2}$
% ...
\end{center}
 
\small
\begin{center}
$^{1}${University of Oklahoma, Norman, Oklahoma 73019}\\
$^{2}${Purdue University, West Lafayette, Indiana 47907}\\
% ...
\end{center}

      Studies of the decays of $B$ mesons to $\psi(2S)$-meson final
      states
      contribute to knowledge of hadronic $B$-meson decays, which involve
      both the weak and strong interactions.  The ARGUS collaboration
      observed the decay $B^+ \to \psi(2S) K^+$~\cite{conj} with a
      branching
      fraction $( 18 \pm 8 \pm 4 ) \times 10^{-4}$ and obtained upper
      limits
      for the branching fractions of the other $B \to \psi(2S) K^{(*)}$
      modes~\cite{argus}.  The CLEO collaboration subsequently measured
      the
      branching fraction $ {\mathcal{B}} (\BMK) = ( 6.1 \pm 2.3
      \pm 0.9 ) \times 10^{-4}$ and determined more stringent upper
      limits for
      the other $B \to \psi(2S) K^{(*)}$ branching
      fractions~\cite{cleopsiold}.  Recently, the CDF collaboration
      measured
      the branching fractions $ {\mathcal{B}} (\BMK) = (5.6 \pm 0.8 \pm
      1.0)
      \times 10^{-4}$ and $ {\mathcal{B}} (\BZKS) = (9.2 \pm 2.0 \pm 1.6)
      \times 10^{-4}$~\cite{cdf}.

      Of the decays $B \to \psi(2S) K^{(*)}$~\cite{kst} reported here,
       the
      modes involving a neutral $B^0$ meson decaying to a $CP$ eigenstate
      can be used, in a manner similar to that for their $J/\psi$
      analogues,
      to measure the $CP$-violation angle $\beta$ of the unitarity
      quark-mixing triangle.  Measurements of the modes $B \to \psi(2S)
      K^{(*)}$ can also contribute to tests of the factorization
      hypothesis~\cite{bsw} and phenomenological techniques 
      employed in several models that
      predict the ratios of vector to pseudoscalar kaon production and
      the
      longitudinal polarization fraction in $B \to J/\psi K^{(*)}$ and $B
      \to \psi(2S) K^{(*)}$ decays
      ~\cite{neubert,deshpande,aleksan,neubert2,chengnew}.  Absolute
      branching fractions have been calculated by combining these
      phenomenological approaches with inputs from
      experiment~\cite{deshpande}.  Nonfactorizable contributions to the
      decay amplitudes can provide substantial corrections to these
      predictions~\cite{nonfactor}.  Both improvements in the accuracy of
      the
      experimental measurements and the observation of new modes can help
      in
      differentiating between models and understanding the role of any
      nonfactorizable corrections~\cite{aleksan,neubert2,chengnew}.

In this Rapid Communication we report measurements of all four decays
$B \to \psi(2S) K^{(*)}$, including the first observation
of the decays $\BZK$ and $\BMSS$. We also present the first angular
analysis of the decays  $\BMSS$ and $\BZSS$, which leads to 
a determination of the longitudinal polarization fraction,
$\Gamma_L/\Gamma$. 
The measurements reported in this Rapid Communication 
supersede the previous CLEO results~\cite{cleopsiold}. 

 The data used in this analysis were collected from $e^+e^-$
collisions on or near the $\Upsilon(4S)$ resonance at the Cornell Electron
Storage Ring (CESR) with two configurations of the CLEO detector, CLEO
II and CLEO II.V.

In CLEO II~\cite{Kub}, the momenta of charged
particles were measured in a tracking system consisting of a
6-layer straw-tube chamber, a 10-layer precision
drift chamber, and a 51-layer main drift chamber, all operating  
inside a 1.5 T solenoidal magnet. The main drift chamber also
provided a measurement of the specific ionization ($dE/dx$) of
charged particles. For CLEO II.V, the innermost wire
chamber was replaced with a three-layer silicon 
vertex detector~\cite{si}, and the argon-ethane gas of the main
drift chamber was replaced with a helium-propane mixture.
A 7800-crystal CsI calorimeter
detected photon candidates and was used for electron
identification. Muon candidates were identified with proportional
counters placed at various depths in the steel absorber.  
The total integrated luminosity of the data sample 
at the $\Upsilon (4S)$ energy is 9.2 fb$^{-1}$,
corresponding to the production of $9.7 \times 10^{6}$ $B\bar{B}$ pairs.
A data sample of 4.6 fb$^{-1}$  recorded 60 MeV below the 
$\Upsilon(4S)$ energy was used for continuum 
non-$B\bar{B}$ background evaluation.
The Monte Carlo simulation of the CLEO detector is
GEANT-based~\cite{geant}. Simulated events for the 
CLEO II and CLEO II.V configurations are processed 
in the same manner as data. 

  Candidates for the decays $B^+ \to \psi(2S) K^{(*)+}$
and $B^0 \to \psi(2S) K^{(*)0}$ are reconstructed via the decays
$\psi(2S) \to \LL$ and 
$\psi(2S) \to J/\psi \pi^+ \pi^- \to \LL \pi^+ \pi^-$,
where $\LL$ stands for 
$e^+e^-$ or $\mu^+ \mu^-$ pairs.
The $K^{*+}$ and $K^{*0}$ mesons are reconstructed in their
$K_S^0 \pi^+$, $K^+ \pi^0$, $K^+\pi^-$, and $K_S^0\pi^0$ modes.

  Electron candidates are identified by their calorimeter energy
deposition, which must be consistent with
their measured momenta and their specific ionization 
in the drift chamber.  Electrons may be accompanied by radiative 
photons emitted in the narrow cone along 
the momentum direction of the electron. 
The recovery of these photons improves the invariant
mass resolution and results in a 20\% relative increase in 
the $\psi(2S) \to \LL$ reconstruction efficiency~\cite{silvia}.
At least one muon candidate is required to penetrate five
nuclear interaction lengths of material, whereas the other
candidate must penetrate at least three nuclear interaction lengths.
In the decays $\psi(2S) \to J/\psi \pi^+ \pi^- $, the
$\pi^+ \pi^-$ invariant mass is required to be greater 
than 0.4 GeV/$c^2$, as motivated by the 
measured $\pi^+ \pi^-$ invariant mass spectrum~\cite{mark}. 
For $J/\psi$ and $\psi(2S)$ candidates in the dielectron final state we use
an asymmetric mass criterion to take into account the radiative tail:
$ -100 < M_{e^+e^-} - M_{J/\psi} < 50$~MeV/$c^2$ and 
$ -140 < M_{e^+e^-} - M_{\psi(2S)} < 60$~MeV/$c^2$.
The dimuon candidate mass is required to be within 50 (60)
MeV/$c^2$ of the $J/\psi$ $\left(\psi(2S)\right)$ mass.

  Candidate $K_S^0$ mesons are reconstructed from pairs of oppositely 
charged tracks with vertices separated from the primary interaction point 
with at least 3 standard deviations. Candidate $K^*$ mesons are 
required to have a $K\pi$ invariant mass within 80 MeV/$c^2$ of the 
 $K^*$ mass~\cite{pdg}. For the charged kaon candidates from $K^*$ decays, 
the $dE/dx$ and time-of-flight information (at least one source
of identification must be available)
must be consistent with a kaon hypothesis to within two standard deviations.

Photon candidates are defined as energy clusters in the calorimeter of
at least 60 MeV in the barrel region, $|\cos\theta|<0.80$, and 100 MeV
in the end cap region, $0.80<|\cos\theta|<0.95$, where $\theta$ is the
polar angle with respect to the beam axis.  Each photon candidate must
have a lateral profile of energy deposition consistent with that
expected of a photon.  In addition, we do not use the fragments of a
nearby large shower as photon candidates.  
The $\pi^0$ candidates are reconstructed from
photon pairs with at least one photon from the barrel region and an
invariant mass within 3 standard deviations 
of the PDG $\pi^0$ mass~\cite{pdg}.  The $\pi^0$ mass
resolution is calculated from the known angular and energy 
resolutions of the calorimeter.

For the modes with a neutral pion in the final state,
the $K^*$ helicity angle 
must be greater than $\pi/2$, which effectively eliminates 
the low momentum neutral pion background. 
The $K^*$ helicity angle, $\theta_{K^*}$,
is the polar angle of the $K$ meson in the $K^*$ rest frame relative to
the negative of the $\psi(2S)$ direction in that frame.

  The $B$ candidates are selected by means of two parameters: the
difference between the energy of the $B$ candidate and the beam
energy, $\Delta E \equiv E(\psi(2S)) + E(K^{(*)}) - E_{\mathrm beam}$,
and the beam-constrained $B$-candidate mass, $M(B) \equiv
\sqrt{E_{\mathrm beam}^2 - \vec{p}_B^2}$, where $\vec{p}_B$ is the
momentum of the $B$ candidate.  The $B$ candidate must be within
the $\pm 3$ standard deviation signal region ( Table~\ref{table:e1} )
in the $\Delta E$ {\it vs.} $M(B)$ plane.

\begin{figure}[hp]
\centering
\epsfxsize=7.5cm
\epsfbox{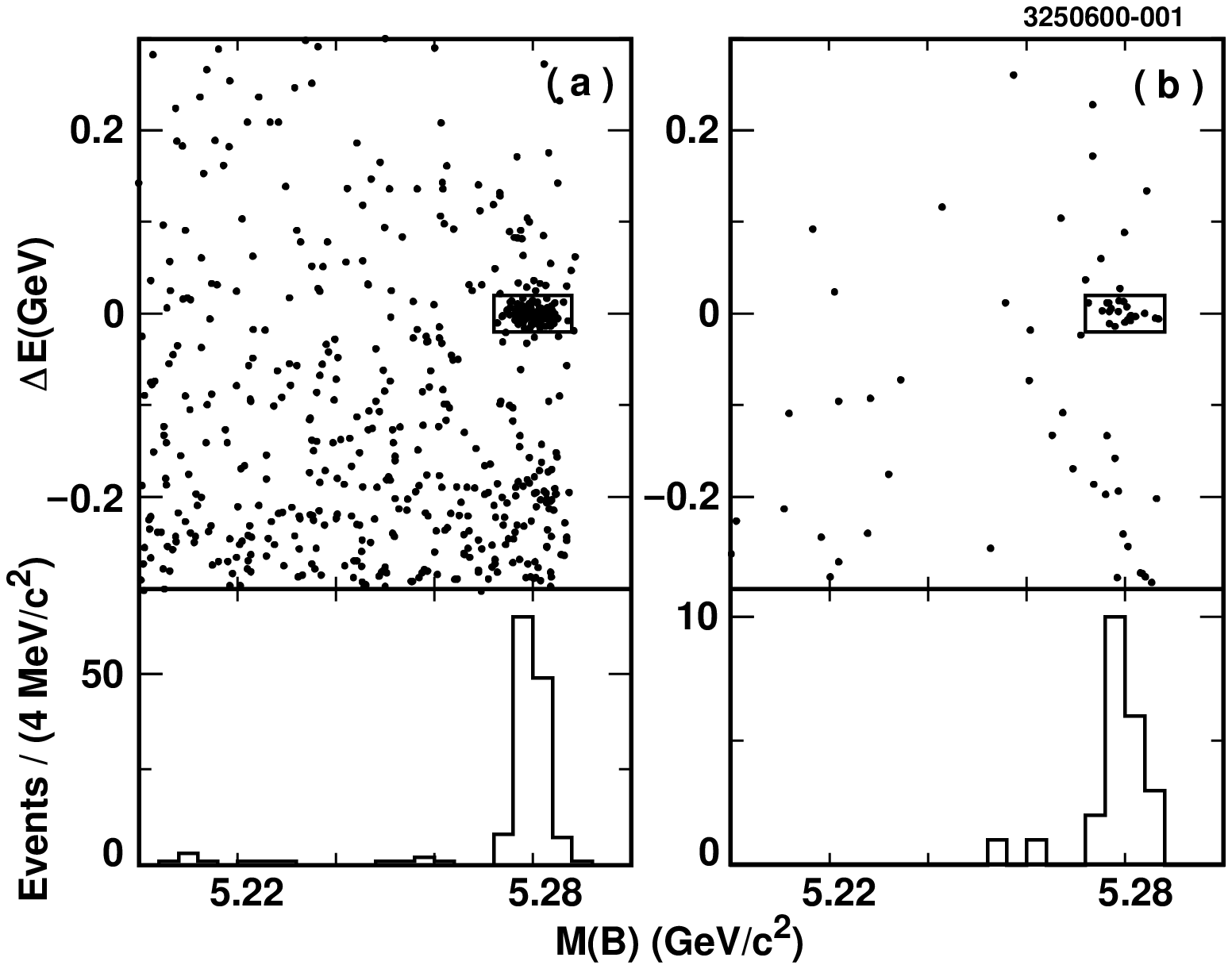}
\epsfxsize=7.5cm
\epsfbox{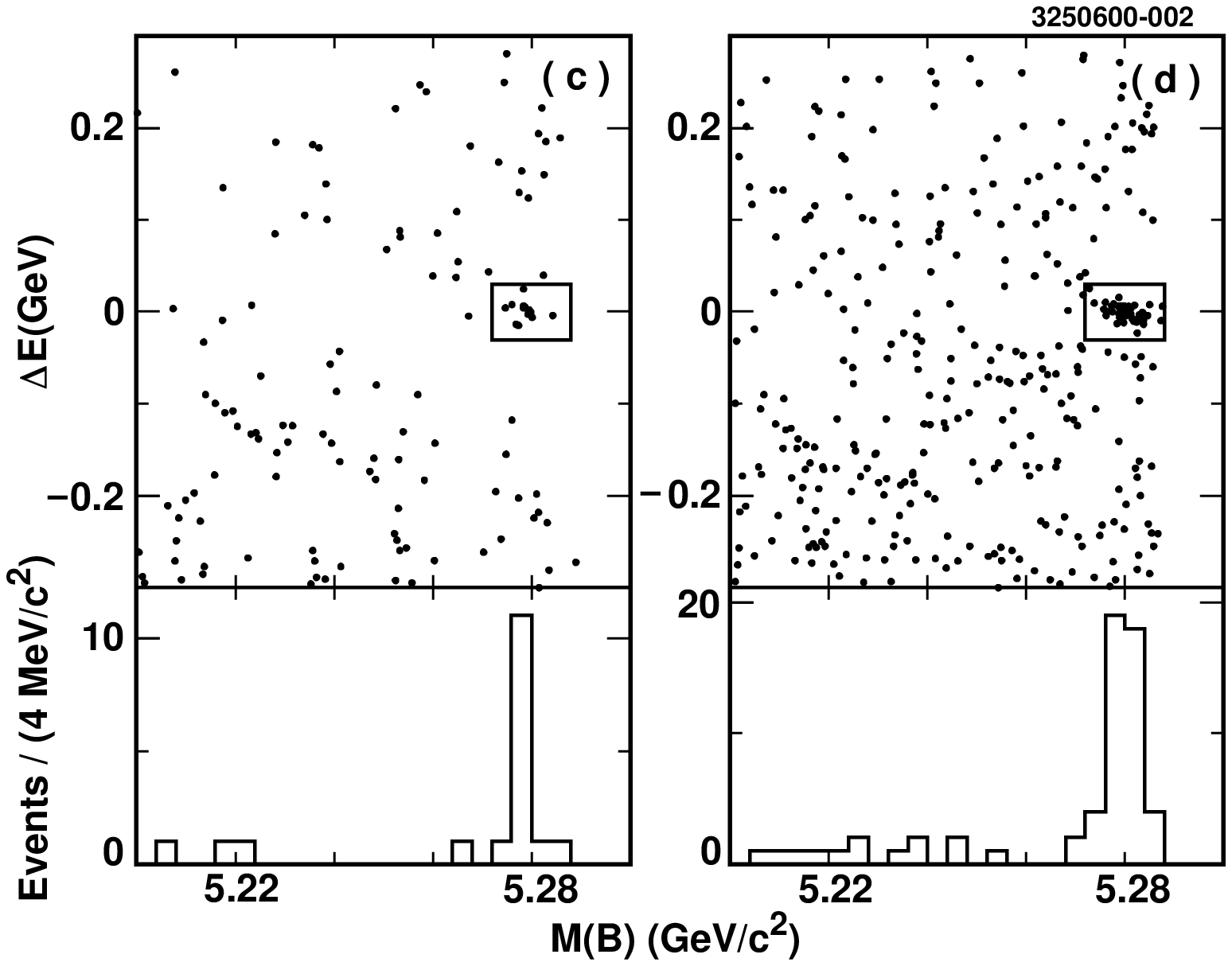}
\epsfxsize=7.5cm
\epsfbox{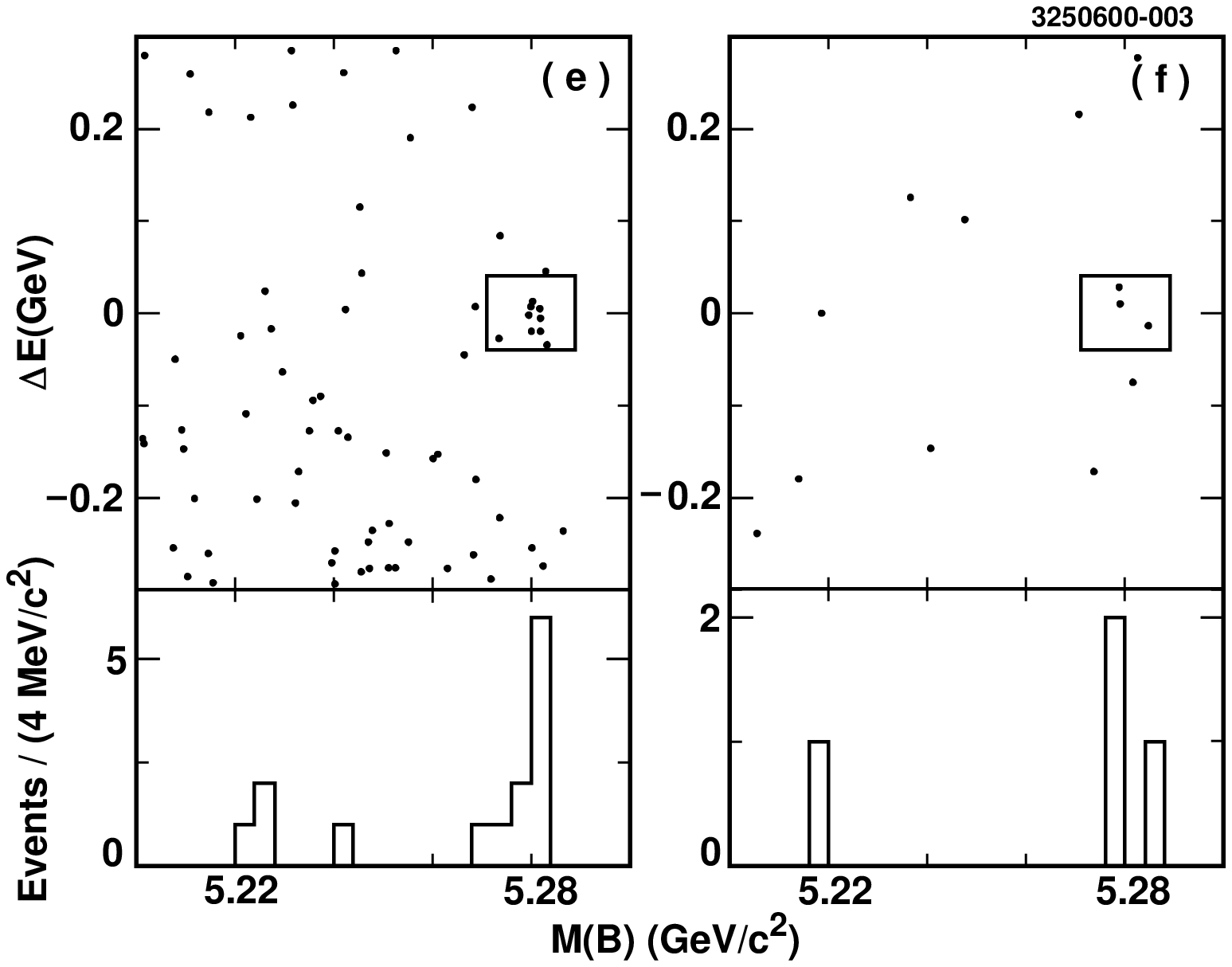}
\caption{$\Delta E$ {\it vs.} $M(B)$ for 			
(a) $B^+ \to \psi(2S) K^+$, 
(b) $B^0 \to \psi(2S) K_S^0$, 
(c) $B^+ \to \psi(2S) K^{*+}$, $K^{*+} \to K_S^0 \pi^+$, 
(d) $B^0 \to \psi(2S) K^{*0}$, $K^{*0} \to K^+ \pi^-$, 
(e) $B^+ \to \psi(2S) K^{*+}$, $K^{*+} \to K^+ \pi^0$, 
(f) $B^0 \to \psi(2S) K^{*0}$, $K^{*0} \to K_S^0 \pi^0$
candidate events, with the contributions from $\psi(2S) \to \LL$ and 
$\psi(2S) \to J/\psi \pi^+ \pi^-$ combined.
The boxes indicate the signal regions.
Also shown are the $M(B)$ projections for the candidate events with
$\Delta E$ within the signal area limits.      
}
\label{fig:sum}
\end{figure}

 After the $B \to \psi(2S) K^*$ event selection, 10 $-$ 20 \% of the
events have more than one $B$ candidate in the signal
area.  In these cases, we select the $B$ candidate with 
minimum $ \Sigma ( x_i - \mu_i )^2 / \sigma_i^2 $,
where $\mu_i$ is a central value of the measured parameter $x_i$ 
and $\sigma_i$ is its uncertainty 
($B \to \LL K^*$ and $B \to \LL \PP K^*$ were considered
different modes).
The following parameters were used where available: the
masses of the $\psi(2S)$, $K^*$, $K_S^0$, and $\pi^0$ candidates, and
the identification significance of the kaon candidates from $K^*$
decays and the pion candidates from the $\psi(2S) \to J/\psi
\pi^+\pi^- $ decay.  The distributions of $\Delta E$ {\it vs.} $M(B)$
for the six different $B \to \psi(2S) K^{(*)}$ decays after 
all selection criteria are applied are shown in
Fig.~\ref{fig:sum}.

   The principal sources of background are misreconstructions of a different
$B \to \psi(2S) K^{(*)}$ mode or $B \to \psi(2S) K \pi \pi $ modes,
combinatorial background from $\Upsilon(4S) \to B\bar{B}$ 
decays that do not contain a $\psi(2S)$ daughter, 
and continuum non-$B\bar{B}$ decays.

   Contributions from miscellaneous $B$ decays with $\psi(2S)$ decay
products are estimated using the Monte Carlo simulation of $B\bar{B}$
events in which one of the $B$ mesons decays exclusively in the
selected mode.  The following modes are considered for calculations of
background from misidentified $B$ decays to states with charmonium: $B
\to \psi(2S) K$ processes with branching fractions obtained in this
Rapid Communication (before correcting for this background);  $B \to
\psi(2S) K^*$ processes with similarly obtained branching fractions
and non-resonant contributions to the $K^*$ reconstruction not
considered; and  $B \to \psi(2S) K \pi \pi$ decays 
with the value of the branching
fraction consisting of that for inclusive $B \to \psi(2S) X$
production~\cite{pdg}, after the subtraction of $K$ and $K^*$ decay
contributions.

  The combinatorial background is estimated with fits of the
beam-constrained $B$ mass distributions in data. The background shape
is obtained with events in the $\Delta E$ sideband areas: $0.05 < |
\Delta E | < 0.15$~GeV.  As a check, the combinatorial background is
also estimated using the $\Upsilon(4S) \to B \bar{B}$ Monte Carlo
sample with $B \to \psi(2S) X $ decays excluded. The results of the
two methods agree within statistical uncertainty.  The results on
signal and background yields are summarized in Table \ref{table:e1}.
Lepton universality is assumed in calculations of the efficiencies for
the $\psi(2S) \to \LL$ mode.

\begin{table}[hp]
\caption{Dimensions of the $\Delta E$ {\it vs.} $M(B)$ signal area 
($M_0$ is the world-average $B$-meson
mass~\protect\cite{pdg}), number of events in the signal area, background
estimates, and detection efficiencies (branching fractions not included).}
\begin{tabular}{lcccccc} 
  &\scriptsize $\BMK$ &\scriptsize $B^0 \to \psi(2S) K_S^0 $ & 
    \multicolumn{2}{c}{\scriptsize $B^+ \to \psi(2S) K^{*+}$} &
    \multicolumn{2}{c}{\scriptsize $B^0 \to \psi(2S) K^{*0}$} \\
  &   &  & 
  \scriptsize $K^{*+} \to K_S^0 \pi^+$ &\scriptsize $K^{*+} \to K^+ \pi^0$ &
  \scriptsize $K^{*0} \to K^+ \pi^-$ & \scriptsize $K^{*0} \to K_S^0 \pi^0$ \\\hline
\footnotesize $| \Delta E |$~~[MeV] & \small 20 & \small 20 & \small 30 & \small 40 & \small 30 & \small 40 \\
\footnotesize $| M(B) - M_0 |$~~[MeV/c$^2$] & \small 8 & \small 8 & \small 8 & \small 9 & \small 8 & \small 9 \\\hline 
\footnotesize $N$($\psi(2S) \to \LL$) & \small 60  & \small 11 & \small 5  & \small 7 & \small 20  & \small 1 \\
\footnotesize $N$($\psi(2S) \to J/\psi \PP$)& \small 69  & \small 10 & \small 9  & \small 2 & \small 25  & \small 2 \\\hline 
\footnotesize $B \to \psi(2S) X $ bkg.   & \small $0.2 \pm 0.1 $ & \small $0.02 \pm 0.02$  & \small $0.6 \pm 0.2 $ & \small $0.3 \pm 0.2$ & \small $1.7 \pm 0.5 $ & \small $0.2 \pm 0.1$ \\
\footnotesize Combinatorial bkg.  & \small $1.6 \pm 0.5$ & \small $ 0.3 \pm 0.2 $ & \small $ 0.5 \pm 0.3 $ & \small $0.7 \pm 0.3 $ & \small $ 1.8 \pm 0.5 $ & \small $0.1 \pm 0.1$ \\
\footnotesize Total bkg.  & \small $1.8 \pm 0.5$ & \small $ 0.3 \pm 0.2 $ & \small $1.1 \pm 0.4$ & \small $ 1.0 \pm 0.4 $ & \small $3.5 \pm 0.7$ & \small $ 0.3 \pm 0.1 $ \\\hline 
\footnotesize $\epsilon$($\psi(2S) \to \LL$)~~[\%] & \small 44 & \small 33 & \small 18  & \small 6 & \small 23 & \small 5 \\ 
\footnotesize $\epsilon$($\psi(2S) \to J/\psi \PP$)~~[\%] & \small 23 & \small 17 & \small 8 & \small 3 & \small 11 & \small 3 \\ 
\end{tabular}
\label{table:e1}
\end{table}

The decays $B \to \psi(2S) K^*$ are a transition from a pseudoscalar to 
a pair of vector mesons. The fraction of longitudinal polarization is 
extracted from the distribution of the $K^*$ helicity angle.
The distribution of the $K^*$ helicity angle is given by~\cite{decay}
$\frac{d\Gamma}{d\cos \theta_{K^*}} \propto 
\sin^2\theta_{K^*} (1 - \Gamma_L/\Gamma) +
2 \cos^2\theta_{K^*} \Gamma_L/\Gamma$.

\begin{figure}[htbp]
\centerline{\epsfig{file=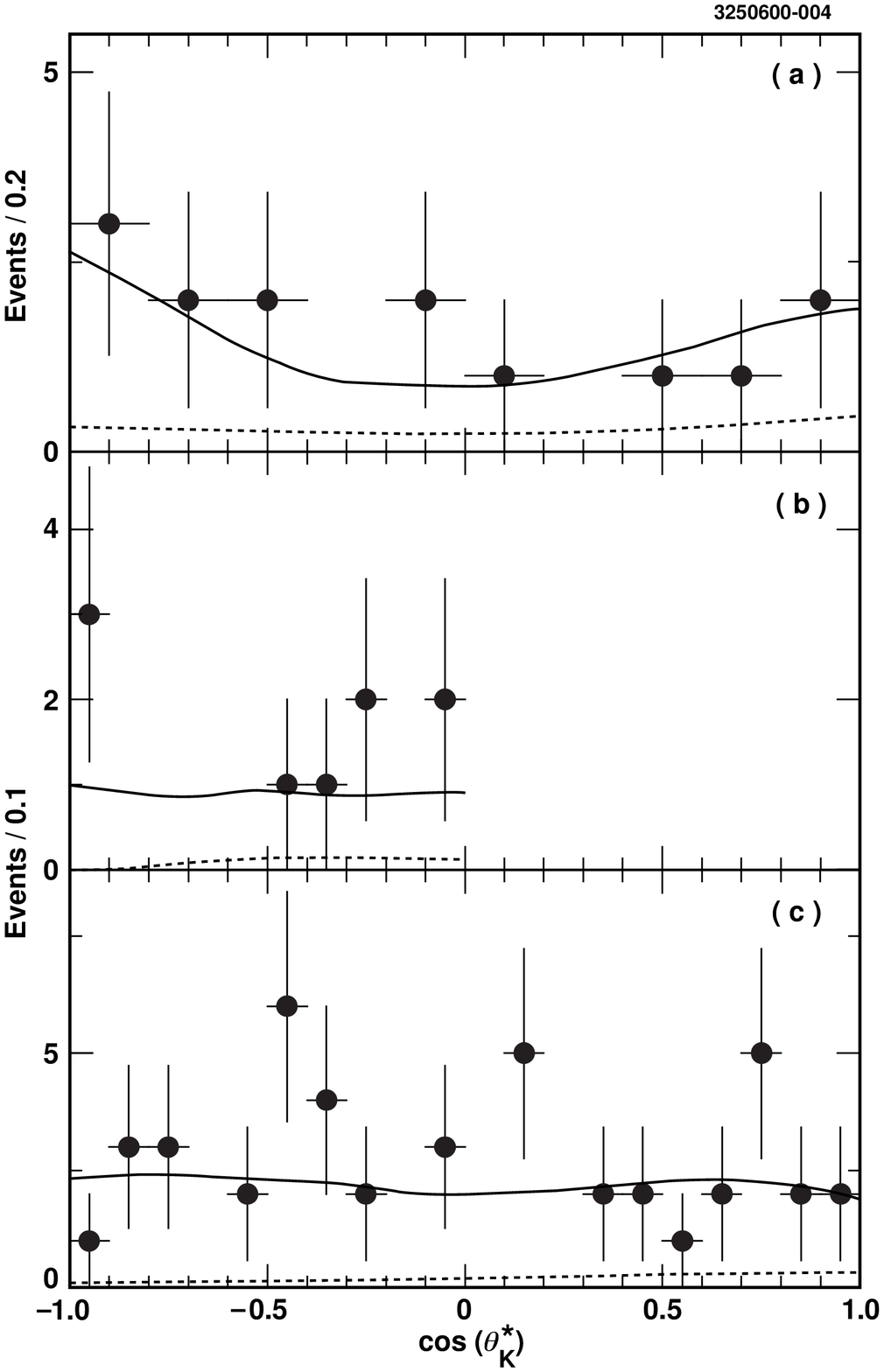,width=2.5in}}
\caption{ Spectra of the $K^*$ helicity angles in   
(a) $B^+ \to \psi(2S) K^{*+}$, $K^{*+} \to K_S^0 \pi^+$;
(b) $B^+ \to \psi(2S) K^{*+}$, $K^{*+} \to K^+ \pi^0$; and 
(c) $B^0 \to \psi(2S) K^{*0}$, $K^{*0} \to K^+ \pi^-$
candidate events in data. The solid curves represent the fit results
to the data (points). The dashed curves represent  
the background contributions.}
\label{fig:cosdat}
\end{figure}

Fig.~\ref{fig:cosdat} shows the  $K^*$ helicity angles for the  
$B^+ \to \psi(2S) K^{*+}$, $K^{*+} \to K_S^0 \pi^+$; 
$B^+ \to \psi(2S) K^{*+}$, $K^{*+} \to K^+ \pi^0$; and 
$B^0 \to \psi(2S) K^{*0}$, $K^{*0} \to K^+ \pi^-$ candidate events in 
data. The $B^0 \to \psi(2S) K^{*0}$, $K^{*0} \to K_S^0 \pi^0$
data are not used in the polarization measurements 
because the lack of statistics precludes a reasonable understanding
of the background.
The curves show the results of the binned likelihood fit to the data.
The fit function 
includes the variable $\Gamma_L/\Gamma$ and a fixed amount of background,
as listed in Table~\ref{table:e1}.
The signal shapes in the fit function for decays with the extreme values
of $\Gamma_L/\Gamma = 0$ and 1 are extracted 
from Monte Carlo simulation. The detector resolution in 
$\cos \theta_{K^*}$ is $\sim$0.06, which is significantly smaller
than the bin width.
The background shape is estimated using the events from sidebands
in the $M(B)$ {\it vs.} $\Delta E$ plane.
The results for the fraction of longitudinal polarization, with
statistical uncertainties only, are 
$0.64 \pm 0.22$, $0.38 \pm 0.23$, and $0.40 \pm 0.14$
for the decays with
$K^{*+} \to K_S^0 \pi^+$, 
$K^{*+} \to K^+ \pi^0$, and 
$K^{*0} \to K^+ \pi^-$ final states, respectively.
The correctness of the fit is checked by fitting 
Monte Carlo generated distributions with a known
value of the longitudinal polarization fraction. The probabilities to get
greater likelihood values than the observed value
are 88, 12, and 10 \% for these $B$ modes, respectively.

 The acceptance and efficiency are evaluated with a simulated
sample of $B \to \psi(2S) K^{(*)}$ decays.
 The contributions to the systematic error come from 
the uncertainty in the reconstruction efficiency 
due to track finding (1\% per track), 
lepton and kaon identification (3\% per candidate),
$K_S^0$ finding (2\% per candidate), $\pi^0$ reconstruction 
(3\% per candidate), 
background evaluation (Table~\ref{table:e1}),
as well as from uncertainties in the $\psi(2S)$ 
and $J/\psi$ branching fractions~\cite{pdg}.
The Monte Carlo statistical uncertainty is at least a factor of 10
smaller than the statistical uncertainty of the data. 
Equal production of charged and neutral $B$-meson pairs in 
$\Upsilon(4S)$ decays is assumed.
In cases of decays $\psi(2S) \to J/\psi \pi^+ \pi^-$,
the additional systematic uncertainty of 2\% comes from the uncertainties 
involved in the generation of the $\pi^+ \pi^-$ invariant mass spectrum.
For the modes with $K^*$ daughters, the efficiency depends on the helicity 
composition of the final state due to the fact that  
the momenta of the $K^*$ decay products
are correlated with the helicity angle. 
The uncertainty in $K^*$ helicity adds a small contribution
of 1\% to the systematic uncertainty (the $\Gamma_L/\Gamma$ result
obtained in this Rapid Communication is used for this estimate).
The major sources of systematic uncertainty in the longitudinal
polarization fraction measurement are
the uncertainties in the fitting procedure (10, 10, 15 \%), background
estimates (5, 15, 5 \%), and differences in detection efficiencies for decays 
with $\Gamma_L/\Gamma = 0$ and 1 (5, 5, 5 \%) for modes with 
$K^{*+} \to K_S^0 \pi^+$, $K^{*+} \to K^+ \pi^0$, and $K^{*0} \to K^+ \pi^-$
final states, respectively.

The results of the  measurements
are summarized in Tables~\ref{table:brres} and \ref{table:polres}.
The branching-fraction results are
$ {\mathcal{B}} (\BMK) = (7.8 \pm 0.7 \pm 0.9) \times 10^{-4}$,
$ {\mathcal{B}} (\BMSS) = (9.2 \pm 1.9 \pm 1.2) \times 10^{-4}$,
$ {\mathcal{B}} (\BZK) = (5.0 \pm 1.1 \pm 0.6) \times 10^{-4}$, and
$ {\mathcal{B}} (\BZSS) = (7.6 \pm 1.1 \pm 1.0) \times 10^{-4}$.
These values supersede the previous CLEO results~\cite{cleopsiold}
and are in agreement with the CDF measurements~\cite{cdf}.
Assuming isospin invariance, we make the first
measurement of the longitudinal polarization 
fraction $\Gamma_L/\Gamma$ in the decays $B \to \psi(2S) K^{*}$,
$\Gamma_L / \Gamma = 0.45 \pm 0.11 \pm 0.04 $, and measure
the ratio of vector to pseudoscalar meson production to be
$ R_{\psi(2S)} \equiv {\mathcal{B}}(B \to \psi(2S) K^*)/ 
{\mathcal{B}}(B \to \psi(2S) K) = 1.29 \pm 0.22 \pm 0.05 $.
Table~\ref{table:compar} compares experimental results 
for $R$ and $\Gamma_L / \Gamma$ with theoretical 
predictions~\cite{neubert,deshpande,neubert2}.
The predictions for $R_{\psi(2S)}$ of Deshpande and Trampetic~\cite{deshpande}
and Neubert and Stech~\cite{neubert2} are inconsistent
with our measurement.

In summary, we have studied all four decays $B \to \psi(2S) K^{(*)}$
with the $\BZK$ and $\BMKS$ modes observed for the first time. 
The first measurement of the longitudinal polarization fraction is 
extracted from an
angular analysis of the $B \to \psi(2S) K^*$ candidates. The
$B^0 \to \psi(2S) K^{(*)0}$ decays are expected to play a significant
role in future CP violation measurements.

We gratefully acknowledge the effort of the CESR staff in providing us with
excellent luminosity and running conditions.
I.P.J. Shipsey thanks the NYI program of the NSF, 
M. Selen thanks the PFF program of the NSF, 
A.H. Mahmood thanks the Texas Advanced Research Program,
M. Selen and H. Yamamoto thank the OJI program of DOE, 
M. Selen and V. Sharma 
thank the A.P. Sloan Foundation, 
M. Selen and V. Sharma thank the Research Corporation, 
F. Blanc thanks the Swiss National Science Foundation, 
and H. Schwarthoff and E. von Toerne
thank the Alexander von Humboldt Stiftung for support.  
This work was supported by the National Science Foundation, the
U.S. Department of Energy, and the Natural Sciences and Engineering Research 
Council of Canada.

\begin{table}[hp]
\caption{Measured branching fractions [$10^{-4}$], where the first
uncertainties are statistical and the second are systematic.
The statistical and uncorrelated systematic uncertainties 
are added in quadrature in calculations of the average
values.}
\begin{tabular}{lc} 
 $ \BMK$ &  $7.8 \pm 0.7 \pm 0.9$ \\
 $ \ \ \ \ \ \BMKS$, $K^{*+} \to K_S^0 \pi^+$ & $8.9 \pm 2.4 \pm 1.2 $ \\
 $ \ \ \ \ \ \BMKS$, $ K^{*+} \to K^+\pi^0$ & $9.8 \pm 3.3 \pm 1.5 $ \\ 
 $ \BMKS$, average & $9.2 \pm 1.9 \pm 1.2$ \\\hline
 $ \BZK$ & $5.0 \pm 1.1 \pm 0.6$ \\
 $ \ \ \ \ \ \BZKS$, $K^{*0} \to K^+ \pi^-$ & $7.5 \pm 1.1 \pm 1.0$ \\
 $ \ \ \ \ \ \BZKS$, $K^{*0} \to K_S^0 \pi^0$ & $12.4 \pm 7.2 \pm 1.8$ \\
 $ \BZKS$, average & $7.6 \pm 1.1 \pm 1.0$ \\
\end{tabular}
\label{table:brres}
\end{table}

\newpage 

\begin{table}[hp]
\caption{Measured longitudinal polarization fractions, $\Gamma_L/\Gamma$,
where the first uncertainties are statistical and the second are systematic.
The statistical and uncorrelated systematic uncertainties 
are added in quadrature in calculations of the average
values.}
\begin{tabular}{lc} 
$ \ \ \ \ \ \BMKS, K^{*+} \to K_S^0 \pi^+$ & $0.64 \pm 0.22 \pm 0.08 $ \\
$ \ \ \ \ \ \BMKS, K^{*+} \to K^+ \pi^0$ & $0.38 \pm 0.23 \pm 0.07 $ \\
$\BMKS$ , average &   $0.51 \pm 0.16 \pm 0.05 $ \\
$\BZKS$ &  $0.40 \pm 0.14 \pm 0.07 $ \\\hline
$ B \to \psi(2S) K^*$, average &  $0.45 \pm 0.11 \pm 0.04 $
\end{tabular}
\label{table:polres}
\end{table}

\begin{table}[hp]
\caption{ Comparison of model predictions and experimental
results for $ R_{\psi(2S)} $ and $\Gamma_L/\Gamma$, where the first
uncertainties are statistical and the second are systematic.}
\begin{tabular}{lcc} 
Source  & $ R_{\psi(2S)} $ & $ \Gamma_L / \Gamma $ \\\hline
Neubert {\it et al.}~\cite{neubert} & 1.85  & - \\
Deshpande and Trampetic~\cite{deshpande} & 3.8  & -  \\
Deandrea {\it et al.}~\cite{deshpande} & 2.0  & - \\
Cheng~\cite{deshpande} & 1.57  & 0.33 \\
Neubert and Stech~\cite{neubert2} & 4.35 & - \\\hline
CDF measurement~\cite{cdf} & $1.62 \pm 0.41 \pm 0.19$ & -  \\
This measurement & $1.29 \pm 0.22 \pm 0.05$ & $ 0.45 \pm 0.11 \pm 0.04 $ \\ 
\end{tabular}
\label{table:compar}
\end{table}



\begin{thebibliography}{99}
\bibitem{conj} Charge conjugation is implied throughout this Rapid 
Communication.
\end{thebibliography}
\end{document}